\documentclass[12pt]{iopart} 
\usepackage{iopams}
\usepackage{graphicx}                                                                                                                        
\begin{document}
\letter{Effect of atomic transfer on the decay of a Bose-Einstein
condensate}
\author{Pawe{\l} Zi\'{n}*\S, Andrzej Dragan*\S, Szymon Charzy\'{n}ski*\ddag,
Norbert Herschbach*, Paul Tol*, Wim Hogervorst* and Wim Vassen*} 

\address{*\ Laser Centre Vrije Universiteit,
De Boelelaan 1081, 1081 HV Amsterdam, The Netherlands}

\address{\S\ Wydzia{\l}
Fizyki, Uniwersytet Warszawski, Ho\.{z}a 69, PL-00-681 Warszawa,
Poland}

\address{\ddag\ Centrum Fizyki
Teoretycznej, Al. Lotnik\'{o}w 32/46 02-668 Warszawa, Poland}

\begin{abstract}
We present a model describing the decay of a Bose-Einstein condensate,
which assumes the system to remain in thermal equilibrium during the
decay. We show that under this assumption transfer of atoms occurs
from the condensate to the thermal cloud enhancing the condensate
decay rate.
\end{abstract}
\pacs{03.75.Hh, 05.30.Jp, 34.50.Fa}
\submitto{\jpb}
\maketitle
\section{Introduction}
Rapid advances in experimental techniques of laser cooling and
trapping made it possible to achieve Bose-Einstein condensation (BEC) 
in weakly interacting
systems. One of the many interesting aspects of BEC in dilute gases is the
dynamics of growth and decay of the
condensate~\cite{Dalibard,Gardiner,Kohl}.
Bose-Einstein condensates can have long lifetimes
ranging from $2$~s~\cite{hel} up to more than
$10$~s~\cite{Dalibard,stampkurn}. This finite lifetime is mainly
caused by inelastic collisions between condensate atoms and by collisions
with particles from the background gas, resulting in atom loss and
heating of the system. Many experiments with condensates can be performed
on a time scale short compared to the condensate lifetime, such
that decay effects are not important. This is, however, not always the
case and some experiments specifically focus on condensate
decay~\cite{Dalibard}.  Hence, it is important to model the decay of a
condensate in detail, taking into account all processes which
significantly contribute. In
literature condensate decay due to inelastic
collisions and collisions with  background gas atoms as well as heating effects
are discussed. To our knowledge,
the transfer of atoms between
the condensate and the thermal cloud occuring dynamically as a
consequence of thermalization of the system has not yet been taken into account. 
We will show that this
process plays an important role when the number of atoms in the
condensate is of the same order or smaller than the number in the
thermal cloud. Its effect could only be neglected in experiments with
large condensate fractions which are often obtained by removal of most
of the thermal part with an rf-knife, or in situations in which our
assumption of fast thermalization is not fulfilled.

The present work concentrates on
the decay of condensates in the presence of a considerably large
thermal fraction. We assume thermal equilibrium during the decay of the
condensate which is justified in most experiments with BEC in dilute
atomic gases. Since elastic collision rates are typically large
($\gtrsim 10^3$~s$^{-1}$~\cite{hel,davis}), one expects thermalization
to occur very rapidly~\cite{snoke} compared to the rate of change in
thermodynamic variables during the decay. Using the simple condensate
growth equation~\cite{Gardiner} we investigated numerically the effect
of dynamical disturbances caused by atom loss and heating from
inelastic collisions on a system originally in thermal equilibrium.
We found that the system stays very close to thermal equilibrium under
conditions typically encountered in the experiments with dilute atomic
gases.

Although our model incorporates two- and three-body collisions it does
not take into account subsequent secondary effects. 
For instance, the effects of induced local
variations in the mean-field interparticle interaction~\cite{Guery-Odelin00}
are neglected. Also secondary collisions of reaction products with
atoms of the condensate or the thermal cloud are not accounted for. 
Avalanches, recently
discussed by Schuster {\em et al}~\cite{Rempe}, are therefore beyond
the scope of this work. They may be incorporated later.

This letter is organized as follows. In Sec.~\ref{need} we derive a
simple analytical formula for the decay of a condensate of
non-interacting bosons including only losses due to collisions with
background gas particles. The expressions derived in this section show the
existence of transfer of atoms between condensate and thermal
cloud and the significant reduction in condensate lifetime that may
occur. In Sec.~\ref{model} we present a more complete dynamical model
of condensate decay using the mean-field theory for weakly-interacting
bosons and including also losses by inelastic two- and three-body
collisions. In Sec.~\ref{results} we present, as an example, numerical
simulations of condensate decay in metastable
helium. 
\section{Atomic transfer}\label{need}
Apart from elastic collisions that keep the system in thermal
equilibrium, in this section we assume the atoms to undergo only collisions with
background gas particles, inducing atom losses. For simplicity we
consider a system of non-interacting bosons. At a temperature $T$ below
the critical temperature $T_c=\hbar\omega[N/g_3(1)]^{1/3}/k$
the number of atoms $N_T$ in the thermal cloud is given
by~\cite{Dalfovo}
\begin{equation}
N_T=g_3(1)\left(\frac{kT}{\hbar \omega}\right)^3,
\label{numthermal}
\end{equation}
where $\omega=(\omega_x\omega_y \omega_z)^{1/3}$ is the averaged
frequency of the harmonic potential trapping the atoms, $N$
denotes the number of atoms and $g_n(u)=\sum_{k=1}^\infty
u^k/k^n$, $g_3(1)\approx 1.202$. The energy $E_T$ contained
in the thermal cloud is~\cite{Dalfovo}
\begin{equation}
E_T=\hbar\omega\frac{\pi^4}{30}\left(\frac{kT}{ \hbar \omega } \right)^4=\alpha N_T^{4/3},
\label{energythermal}
\end{equation}
with $\alpha=\pi^4 \hbar \omega g_3^{-4/3}(1)/30$. To describe the
equilibrium state we choose instead of the temperature $T$ and the
total number of atoms $N$ as dynamical variables the number of atoms
in the condensate $N_C$ and the number of atoms in the
thermal cloud $N_T$.

For the background gas particles we assume room-temperature energies
and a uniform distribution over the volume occupied by the trapped
atoms. As the average kinetic energy of a background gas particle is
much larger than the energy of an atom in the trap, the collision
cross section does not depend on the latter and each collision removes
one atom from the trap. After such a collision the total energy of the
system $E$ is depleted on average by the mean energy per atom $E/N$.
With these assumptions we obtain for the loss rate 
\begin{equation}
\dot{N}=-\frac{1}{\tau}N=\dot{N}_C+\dot{N}_T=-\frac{1}{\tau}(N_C+N_T),
\label{lossrate1}
\end{equation}
with $\tau$ denoting the trap lifetime. For the rate of change in
the total energy $E$ of the system accompanying the atom loss we have
\begin{equation}
\dot{E}=-\frac{1}{\tau}E=\dot{E}_C+\dot{E}_T=-\frac{1}{\tau}(E_C+E_T),
\label{energychange1}
\end{equation}
where we used the energy contained in the condensate
$E_C=\varepsilon_0 N_C$ with
$\varepsilon_0=\frac{1}{2}\hbar(\omega_x+\omega_y+\omega_z)$ being the
ground state energy of the trap. Using Eq.~(\ref{energythermal}) the
time derivatives $\dot{E}_T=\frac{4}{3}\alpha
N_T^{1/3}\dot{N}_T$ and
$\dot{E}_C=\varepsilon_0\dot{N}_C$ can be
substituted in Eq.~(\ref{energychange1}) and in combination with
Eq.~(\ref{lossrate1}) we then find
\begin{equation}
\dot{N}_T = - \frac{1}{\tau} \, N_T \, \frac{ 1 - \varepsilon_0 N_T/ E_T}{4/3 -\varepsilon_0 N_T / E_T}\simeq-
\frac{3}{4 \tau}N_T,
\end{equation}
where we used $\varepsilon_0 N_T/E_T\ll 1$, neglecting 
the energy of a condensate atom compared to the average energy per
atom in the thermal cloud. Substituting in Eq.~(\ref{lossrate1}) yields
\begin{equation}
\dot{N}_C = - \frac{1}{\tau} \left( N_C+
\frac{1}{4} N_T \right).
\label{becdec1}
\end{equation}
This simple analysis shows that for a condensate coexisting in thermal
equilibrium with a thermal cloud in a trap neither
$\dot{N}_C=-N_C/\tau$ nor $\dot{N}_T=
-N_T/\tau$ holds for the decay induced by collisions with
background particles. Conservation of energy and number of atoms
combined with rapid thermalization inevitably results in transfer
of atoms from condensate to thermal cloud thereby enhancing the decay rate of
the condensate. Especially with a considerable fraction of thermal
atoms in the system this affects the decay of the condensate. Only in the
limit of a large condensate fraction Eq.~(\ref{becdec1}) becomes
$\dot{N}_C=-N_C/\tau$.
\section{Interacting model}\label{model}
In order to parametrize the equilibrium state of the gas, we will use
the temperature $T$ and the number of atoms in the condensate
$N_C$ as independent variables fully describing the state of
the system. We will start with the stationary ``two-gas'' model
proposed by Dodd {\it et al}~\cite{Dodd}, in which atoms of the thermal cloud do
not affect the condensate described by the stationary Gross-Pitaevskii
(GP) equation. The thermal cloud atoms do not interact with each other
(except for thermalization) but they are influenced by the condensate
through the mean-field potential $2 U_0 n_C({\bf r})$. Here, the
contact potential $U_0=4 \pi \hbar^2 a/m$ expresses the binary
interaction between atoms with scattering length $a$ and mass $m$, and
$n_C({\bf r})$ denotes the spatial density of the condensate
which can be calculated from the GP equation. We additionally simplify
the two-gas model by describing the thermal cloud semiclassically,
replacing discrete states when evaluating statistical averages by a
continuum of states. This way we obtain analytical formulas for the
atomic density $n_T({\bf r})$ \cite{Dalibard} and energy
density $e_T({\bf r})$ in the thermal cloud:
\begin{eqnarray}
\label{nt} n_T({\bf r})&=&\int \frac{d^3 p}{(2\pi\hbar)^3}
\frac{1}{\exp\left[\left(\frac{p^2}{2m}+V_{eff}({\bf r})-\mu
\right)/kT\right]-1} \nonumber \\ &=&\lambda^{-3}
g_{3/2}\left(e^{-\left(V_{eff}({\bf r})-\mu\right)/kT}\right),
\end{eqnarray}
\begin{eqnarray}
\label{et} e_T({\bf r})&=& \int \frac{d^3 p}{(2\pi\hbar)^3} \frac{\frac{p^2}{2m}+V_{eff}({\bf r})}
{\exp\left[\left(\frac{p^2}{2m}+V_{eff}({\bf r})-\mu \right)/kT
\right]-1} \nonumber \\ \nonumber \\ &=&\frac{3}{2} kT \lambda^{-3}
g_{5/2}\left(e^{-\left(V_{eff}({\bf r})-\mu\right)/kT}\right) +
\nonumber \\ & & V_{eff}({\bf r})\,\lambda^{-3} g_{3/2}\left(
e^{-\left(V_{eff}({\bf r})-\mu\right)/kT}\right),
\end{eqnarray}
where $\lambda=\sqrt{2\pi\hbar^2/mkT}$ \ is the thermal de Broglie
wavelength, $\mu$ is the chemical potential and \ $V_{eff}({\bf
r}) = V({\bf r})+ 2 U_0 n_C({\bf r}) $. The explicit form of
$n_C({\bf r})$ and $\mu$ follows from the GP equation. They
depend on the number of atoms in the condensate $N_C$. By
integrating the densities $n_C$ and $n_T$ over
spatial degrees of freedom we obtain the number of atoms in the
thermal cloud $N_T(N_C,T )$ and the energy of the
thermal cloud $E_T(N_C,T)$.

We are interested in the time dependence of $N_C$ and $T$. As
before, we will first consider the dynamics of the total number of
trapped atoms $N$ and total energy $E$ as a function of $N_C$
and $T$. The advantage of this approach is that we do not need to
state transfer terms explicitly; the transfer will follow from our
analysis automatically.

The total loss rates $\dot{N}$ and $\dot{E}$ are related to
$\dot{N}_C$ and $\dot{T}$ via
\begin{eqnarray}
\label{repa} \dot{N} &=& \frac{\partial N_T }{\partial T} \dot{T} +
\left(\frac{\partial N_T }{\partial N_C} + 1\right)
\dot{N}_C \nonumber \\ \\ \nonumber \dot{E} &=&
\frac{\partial E_T }{\partial T} \dot{T}
+\left(\frac{\partial E_T }{\partial N_C} + \mu
\right) \dot{N}_C,
\end{eqnarray}
with $\mu=\partial E_C/\partial N_C$. The
reparametrization is straightforward as we have already found
explicit expressions of almost all terms appearing in
Eq.~(\ref{repa}).

In order to describe the dynamics of the system we must consider all
relevant processes affecting its state. We include three main
dynamical effects that may cause losses of atoms from the system:
two-body inelastic collisions, three-body recombinations and
collisions with background gas. In the following we will neglect all
secondary collisions. Then the total atomic loss rate $\dot{N}$ is
given by \cite{Kagan}:
\begin{eqnarray}
\label{L} -\dot{N} &=& \frac{1}{\tau} N + 2\chi \int \mbox{d}^3 r\, \left(
\frac{1}{2!} n_C^2+ 2n_C n_T+
n_T^2 \right)+ \nonumber \\ \nonumber \\ & & 3\xi \int
\mbox{d}^3 r\, \left( \frac{1}{3!} n_C^3 + \frac{3}{2!}
n_C^2 n_T + 3n_C n_T^2 +
n_T^3 \right),
\end{eqnarray}
where $\chi$ and $\xi$ are two- and three-body collision rate
constants and $\tau$ is the lifetime of the trap.

Each term in Eq.~(\ref{L}) corresponds to a loss process that may
occur in the system. For example the term
$\xi\frac{3}{2!}n_C^2({\bf r}) n_T({\bf r})$
expresses the probability density of a three-body recombination
between two condensate and one thermal cloud atom to take place at
point ${\bf r}$. In this event we lose all three atoms from the
system, hence the factor $3$ in front of the integral in
Eq.~(\ref{L}). The energy lost in this particular process consists of
two terms. First, an energy $2\mu$ carried by the two lost condensate
atoms. Second, the energy carried by the thermal cloud atom just
before the recombination. We assume that the change in internal energy
per lost thermal particle is equal to the average energy per atom
$e_T({\bf r})/n_T({\bf r})$ at point ${\bf
r}$. Then the energy loss rate by this process becomes
\begin{equation}
\xi\int \mbox{d}^3 r\,\frac{3}{2!}n_C^2
n_T\left(2\mu+\frac{e_T}{n_T}\right).
\end{equation}
Analogously, one can derive the proper rates corresponding to the
other possible collisions. Finally we end up with the expression for
the total energy loss rate:
\begin{eqnarray}
\label{Lfalka} -\dot{E} &=& \frac{1}{\tau}(E_T+\mu N_C)
+ 2\chi \int \mbox{d}^3 r\, \left[ \frac{e_T} {n_T}
\left( n_T^2+ n_Cn_T \right)\right. +
\nonumber \\ \nonumber \\ & & \left.\mu\left(\frac{1}{2!}
n_C^2 +n_Cn_T\right) \right] + \nonumber \\
\nonumber \\ & & 3 \xi \int \mbox{d}^3 r\, \left[ \frac{e_T}
{n_T} \left( \frac{1}{2!} n_C^2 n_T + 2
n_C n_T^2 + n_T^3 \right) + \right.
\nonumber \\ \nonumber \\ & &\left.\mu \left(\frac{1}{3!}
n_C^3+\frac{2}{2!}  n_C^2 n_T +
n_C n_T^2 \right) \right].
\end{eqnarray}
In equations (\ref{repa}), (\ref{L}), and (\ref{Lfalka}) $\dot{N}$,
$\dot{E}$, $N_T$, $E_T$, and $\mu$ are functions of
$N_C$ and $T$ only.
\section{Results and conclusions} \label{results}
\begin{figure}
\begin{center}
\includegraphics[width=0.60\linewidth]{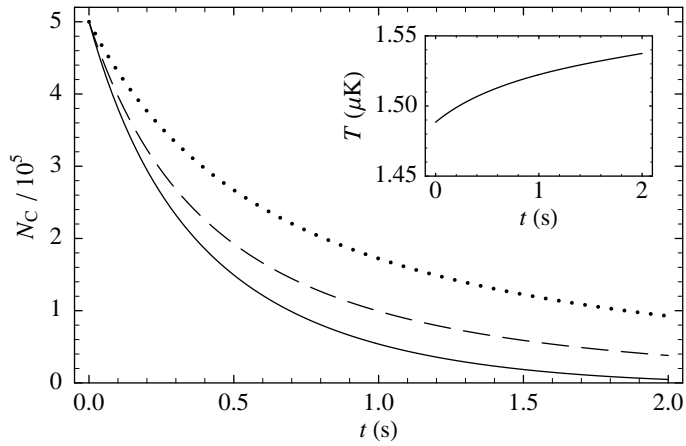}
\caption{\label{Fig:w1} Decay of a condensate of metastable helium atoms with
$N_C(0)=5\times 10^5$ applying our
model (solid line) in comparison with a simpler model neglecting
thermalization (dashed line). The initial number of thermal
atoms $N_T(0)=5\times 10^5$
[$T(0)\approx1.5$~$\mu$K]. The dotted line represents the decay of a pure condensate:
$N_T(0)=0$ ($T=0$~K). Inset: dynamics of
the temperature in our model.}
\end{center}
\end{figure}

\begin{figure}
\begin{center}
\includegraphics[width=0.60\linewidth]{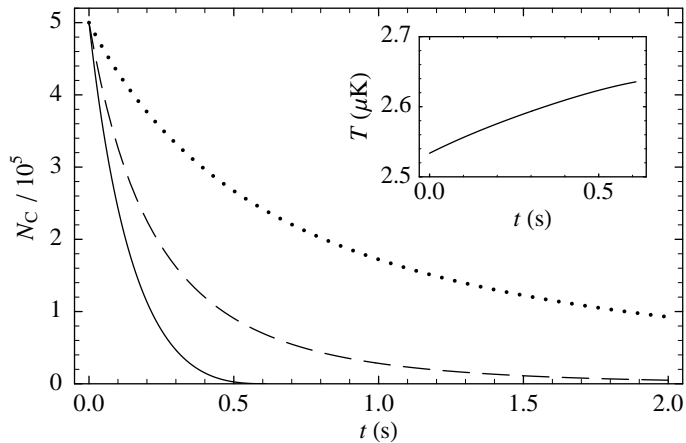}
\caption{\label{Fig:w3} Decay of a condensate with a
large thermal cloud: $N_T(0)=2\times 10^6$
[$T(0)\approx2.5$~$\mu$K]. Further details as in the caption of Fig.~\ref{Fig:w1}.}
\end{center}
\end{figure}

In this section we present some numerical solutions of the set of
equations (\ref{repa}), (\ref{L}), and (\ref{Lfalka}). We have
performed simulations of the condensate decay for the conditions of
the Paris experiment on metastable helium \cite{hel}: an initial
number of atoms in the condensate of $N_C(0)=5\times 10^5$, a
trap lifetime $\tau=35$~s, s-wave scattering length $a=16$~nm, trap
parameters $\omega_x=\omega_y=1090$~Hz and $\omega_z=115$~Hz, two-body
inelastic collision rate $\chi=1.5\times
10^{-14}$~cm$^3/s$ \cite{Shlyapnikov1}, and three-body
recombination rate $\xi=4\times 10^{-27}$~cm$^6/s$
\cite{Shlyapnikov2}. The initial number in the thermal cloud is either
$N_T(0)=5\times 10^5$ (Fig.~\ref{Fig:w1}) or
$N_T(0)=2\times 10^6$ (Fig.~\ref{Fig:w3}), with initial
temperatures of $1.5$~$\mu$K and $2.5$~$\mu$K,
respectively. This gives densities still outside the regime where
avalanches occur. The numerical solution $N_C(t)$ is shown
together with results for a much simpler approach in which the
transfer is absent. For the loss rates $\dot{N}_C$ and
$\dot{N}_T$ again Eq.~(\ref{L}) is used, setting to zero the
thermal or condensate density, respectively.
Comparing both curves with the same initial conditions it follows
that, when the existence of transfer is neglected, the lifetime of the
trap is overestimated by $50\%$ when the initial fraction of atoms in
the thermal cloud is 0.5 and by more than a factor of 2 when this
fraction is 0.8. The third (dotted) line in each figure shows the
decay in the absence of a thermal cloud. The lifetime increases
by a factor of 2.5 in Fig.~\ref{Fig:w1} and a factor of 8 in
Fig.~\ref{Fig:w3}, due to the absence of inelastic collisions between
condensate and thermal cloud atoms. In both models the decay rate is
larger with a thermal cloud, but in the full model the effect is
enhanced by atomic transfer.

Another feature of condensate decay that can be studied on the basis
of our model is the time dependence of the temperature during the
decay. Usually it is assumed that the temperature is constant
\cite{Dalibard}. In both figures an inset shows the numerical solution
for $T(t)$. Indeed, the temperature remains constant within $4\%$.

We have presented an equilibrium model of the decay of a Bose-Einstein
condensate. Our analysis has shown that the assumption of sustained
thermal equilibrium leads to transfer of atoms from the condensate to
the thermal cloud which can significantly enhance the condensate decay
rate. This effect could be seen by an experimental examination of the
decay rate as a function of the fraction of thermal atoms.
\section*{Acknowledgements}
This work was done as part of a student project within the
Socrates-Erasmus exchange programme 2000/2001 at the
Vrije Universiteit in
Amsterdam. P. Z. acknowledges the support from Polish KBN grant
2/PO3/BO7819. We would like to thank Marek Trippenbach and Kazimierz
Rzazewski for useful discussions.
\end{document}